\documentclass[manuscript,authorversion,nonacm]{acmart}

\AtBeginDocument{%
  }

\usepackage{xcolor}
\usepackage{subcaption}
\usepackage{hyperref}
\usepackage{float}
\usepackage{booktabs}
\usepackage{multirow}
\usepackage{cleveref}[2012/02/15]
\crefformat{footnote}{#2\footnotemark[#1]#3}

\acmISBN{978-1-4503-XXXX-X/2018/06}

\begin{document}

\title{MetaCues: Enabling Critical Engagement with Generative AI for Information Seeking and Sensemaking}

\author{Anjali Singh}
\email{anjali.singh@ischool.utexas.edu}
\affiliation{%
  \institution{The University of Texas at Austin}
  \country{USA}
}

\author{Karan Taneja}
\email{ktaneja6@gatech.edu}
\affiliation{%
  \institution{Georgia Institute of Technology}
  \country{USA}
}

\author{Zhitong Guan}
\email{klarazt@utexas.edu}
\affiliation{%
  \institution{The University of Texas at Austin}
  \country{USA}
}

\author{Soo Young Rieh}
\email{rieh@ischool.utexas.edu}
\affiliation{%
  \institution{The University of Texas at Austin}
  \country{USA}
}

\renewcommand{\shortauthors}{Trovato et al.}

\begin{abstract}
Generative AI (GenAI) search tools are increasingly used for information seeking, yet their design tends to encourage cognitive offloading, which may lead to passive engagement, selective attention, and informational homogenization. Effective use requires metacognitive engagement to craft good prompts, verify AI outputs, and critically engage with information. We developed \textit{MetaCues}, a novel GenAI-based interactive tool for information seeking that delivers \textit{metacognitive cues} alongside AI responses and a note-taking interface to guide users' search and associated learning. Through an online study ($N=146$), we compared MetaCues to a baseline tool without cues, across two broad search topics that required participants to explore diverse perspectives in order to make informed judgments. Preliminary findings regarding participants' search behavior show that MetaCues leads to increased confidence in attitudinal judgments about the search topic as well as broader inquiry, with the latter effect emerging primarily for the topic that was less controversial and with which participants had relatively less familiarity. Accordingly, we outline directions for future qualitative exploration of search interactions and inquiry patterns.
\end{abstract}


\settopmatter{printacmref=false}
\setcopyright{none}
\renewcommand\footnotetextcopyrightpermission[1]{}
\pagestyle{plain}
\maketitle

\section{Introduction}
Generative AI (GenAI) search tools, such as Perplexity.ai and Google AI Overview, have been rapidly adopted and promoted by major tech companies. These tools are typically valued for their speed, convenience, and the ability to synthesize information from multiple relevant sources into coherent responses \cite{zhou2024understanding}. 
However, the growing use GenAI for information seeking has raised several concerns. The design of GenAI search tools inherently encourages cognitive offloading \cite{singh2025protecting}, which refers to the delegation of cognitive processes to an external system, such as AI \cite{risko2016cognitive}. In particular, using GenAI for information seeking can bypass the cognitive processes of understanding, applying, analyzing, and evaluating information, potentially undermining learning and critical engagement with information \cite{narayanan2025search, singh2025protecting}.
Research suggests that information seeking with GenAI can lead to passive engagement and limit exposure to diverse perspectives, which can contribute to informational homogenization \cite{amer2024end, narayanan2025search, solaiman2023evaluating}. Beyond influencing search behavior, the fluent, confident, and sycophantic nature of GenAI responses can also shape users' confidence in the judgments they form based on AI-generated content \cite{amer2024end, singh2025protecting}. 
These concerns are particularly salient in the context of controversial topics, where people are more likely to align with information that reinforces their pre-existing beliefs \cite{vedejova2022confirmation}. In such settings, the tendency of GenAI systems to exhibit sycophancy may further amplify confirmation bias.

Given these challenges, effective use of GenAI tools for information seeking requires \textit{metacognitive engagement}, which involves being aware of and regulating over one’s thinking \cite{winne2017cognition, tankelevitch2024metacognitive}. 
Metacognition---commonly referred to as ``thinking about thinking''---is essential not only for effectively prompting GenAI, which requires monitoring task goals and planning, but also for being vigilant of potential hallucinations in AI-generated responses \cite{tankelevitch2024metacognitive}.
Recent work \cite{singh2025enhancing} shows that metacognitive cues---that prompt users to pause, reflect, assess their comprehension and consider multiple perspectives---delivered while searching with GenAI tools can foster active engagement, broader exploration, and more thoughtful follow-up questioning. However, since that study employed a Wizard-of-Oz setup, where researchers manually delivered cues while monitoring participants' search behavior, it remains unclear whether such support can be provided autonomously.

\begin{figure}[t]
  \centering  \includegraphics[width=\columnwidth]{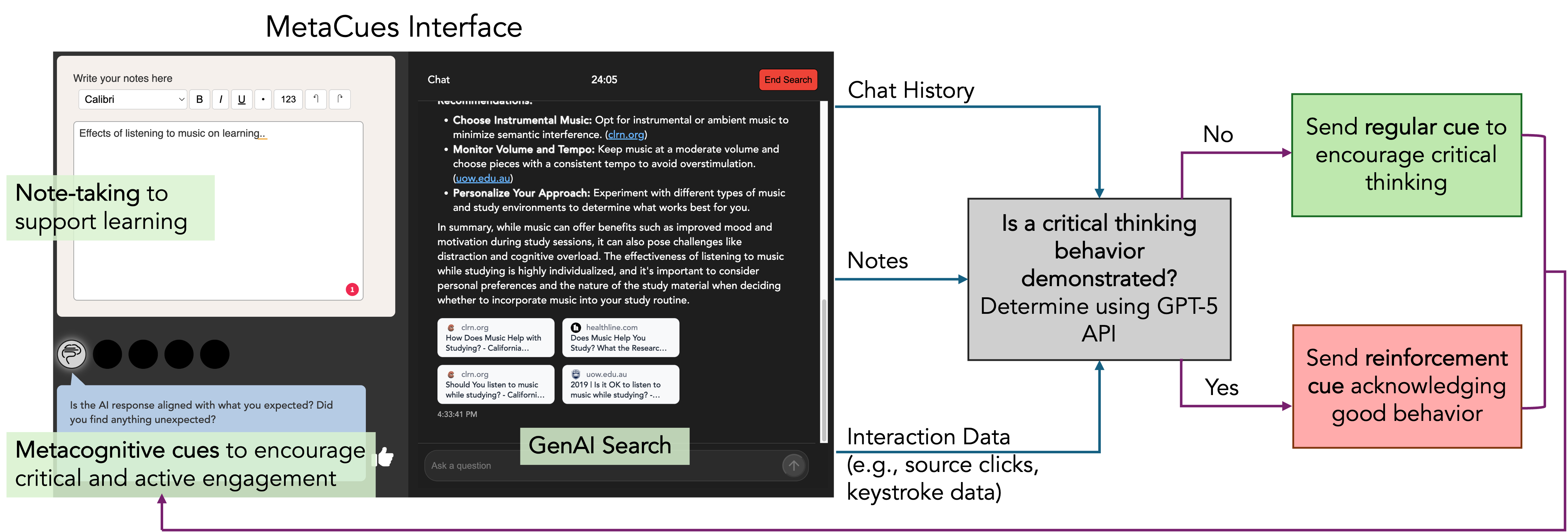}
  \caption{A snapshot of the MetaCues interface, and the process of generating metacognitive cues based on user-generated data.} 
    \label{fig:metacues}
    \vspace{-4mm}
\end{figure}

To address this gap, we designed and developed \textbf{MetaCues} (Figure \ref{fig:metacues}), an interactive GenAI-based tool to support information seeking featuring three panels: a chat interface on the right where users query the AI and view responses with linked sources, and a notepad on the left, below which automatically generated metacognitive cues are displayed. MetaCues analyzes users' questions, AI responses, and notes to proactively deliver tailored cues that encourage active and critical user engagement during the information seeking process.

To examine the effects of metacognitive cues on users’ search behavior and resulting judgments across different search topics, we conducted an online study (N=146) comparing MetaCues to a baseline system, which was identical in every respect except for the absence of cues. 
Participants were randomly assigned one of two information-seeking tasks.
This work describes the design and development of the MetaCues system and reports preliminary findings from this study addressing the following research question: \textit{What are the effects of metacognitive cues on: (i) users' search behavior, and (ii) confidence in their attitudinal judgments regarding the assigned topic, during a GenAI-assisted information-seeking and sensemaking task, compared to not receiving any cues?}

\section{Related Work}
The integration of GenAI into search represents a shift from query– document matching to generating synthesized, context-aware responses \cite{trippasReportFourthStrategic2025, zhaiLargeLanguageModels2024}. 
Generative Information Retrieval systems go beyond ranking results to summarizing, comparing, and explaining information, enabling conversational and adaptive user engagement. 
GenAI search tools promise efficiency and interactivity by generating synthesized, context-aware responses, and supporting conversational and adaptive user engagement \cite{trippasReportFourthStrategic2025, zhaiLargeLanguageModels2024}. However, the fluent and confident nature of their responses can obscure inaccuracies and biases, which can cause overtrust and misinformation \cite{amer2024end, kaiser2025new}. Over-reliance on GenAI tools can cause cognitive offloading, reducing users’ active engagement and critical evaluation of information \cite{risko2016cognitive, fanBewareMetacognitiveLaziness2025, lee2025criticalthinking}. Consequently, GenAI interactions have been found to reinforce existing beliefs \cite{sharma2024generative}
and marginalize alternative viewpoints \cite{amer2024end}.
Given these concerns, the design of GenAI systems imposes metacognitive demands on users \cite{tankelevitch2024metacognitive} and effective interaction with GenAI systems requires metacognitive engagement \cite{flavell1979metacognition, winne2017cognition}. 
For GenAI-assisted information seeking, effective evaluation of AI responses requires users to accurately judge their topic knowledge and adapt prompting strategies as needed \cite{tankelevitch2024metacognitive}. However, this can be challenging for users who have misplaced confidence in their knowledge of a topic or prompting abilities \cite{singh2025enhancing, tankelevitch2024metacognitive}. 

Recent work shows \textit{metacognitive cues} can significantly support information-seeking and sensemaking by promoting reflection, broadening exploration, and deepening inquiry \cite{singh2025enhancing}. Metacognitive cues draw on the concept of metacognitive prompting, an established instructional approach that has been found to support metacognition in educational contexts \cite{bannert2013scaffolding, lin1999supporting} and improve information seeking with traditional search engines \cite{hwang2011information, stadtler2008effects}. Metacognitive cues are designed to direct people’s attention toward their own thought processes and toward understanding the activities in which they are engaged \cite{lin2001designing}. 
Such cues are typically framed as thoughtful questions that are intended to support people’s monitoring and control of their information processing by inducing metacognitive and regulative activities, such as orientation, goal specification, planning, monitoring, and control as well as evaluation strategies \cite{bannert2013scaffolding}.
Their use has been shown to significantly enhance the effectiveness of information seeking in traditional search processes \cite{zhou2019metacognitive}. More recently, emerging GenAI tools have begun to incorporate reflective scaffolds \cite{duelenSocraticAIDisinformation2024, gmeiner2025metacognitive} and Socratic dialogue mechanisms \cite{faveroEnhancingCriticalThinking2024} to increase users’ awareness of their reasoning processes, promote critical engagement with information, and support the detection of misinformation. Building on this body of work, the present study investigates the automatic generation of metacognitive cues within \textit{MetaCues}, a novel GenAI-based information seeking tool, and examines their effects on users’ search behaviors during exploratory sensemaking tasks.

\section{MetaCues System Design}

MetaCues is designed for GenAI-based information seeking with metacognitive guidance, and is served as a web application (Figure \ref{fig:metacues}). 
The application acts as a learning companion that not only provides answers to user queries but also guides the sensemaking process through metacognitive cues that are generated based on user's chat history, notes and click-stream data.

\subsubsection*{\textbf{Chat Interactions and Search}}

The conversational AI interface uses OpenAI GPT-4o model with web search capabilities. 
Temperature is set to 0.8 to balance focus and diversity. 
Search context country is set to the U.S., and search context size is kept low to minimize response time. 
The chat uses an instructional LLM prompt\footnote{\label{main_footnote} \href{https://osf.io/zghqw/overview?view_only=659452f1dce1493780b1ee5b64479a1a}{Supplementary material with LLM prompts and Cue Messages}} that defines the role of AI as a teaching assistant for providing comprehensive academic information. 
We prompt GPT to: (i) provide responses with clarity, conciseness, and minimal jargon, (ii) mandatorily use information from web search with 5+ sources and include citations, (iii) frame the response at a technical level of a Bachelor's degree student, and (iv) structure the response in alignment with major themes in a Markdown format with headings, lists, and emphasis. 
We also prompt the model to respond with `Sorry I can't help you with that' for off-topic queries to promote safety.
Additionally, responses include visual link cards at the end to provide a compact summary of sources provided in the response. 

\subsubsection*{\textbf{Cue Generation Process}}

The cues implemented in MetaCues were informed by prior work by \citet{singh2025enhancing} on metacognitive support in GenAI-based search. Building on insights regarding the metacognitive demands of GenAI tools \cite{tankelevitch2024metacognitive}, GenAI-assisted search \cite{sharma2024generative, narayanan2025search}, and search as a learning process \cite{rieh2016towards}, Singh et al. initially proposed five types of metacognitive cues: Orienting, Monitoring, Comprehension, Broadening Perspectives, and Consolidation. Each cue was delivered according to predefined criteria based on observable user behavior. Following data collection, their study identified measurable indicators of critical thinking, called Persistent Inquiry, Independent Thinking, and Source Engagement, and recommended tailoring cues to support these behaviors.

Guided by these insights, we adopted four cue types directly from Singh et al.---\textit{Orienting, Monitoring, Broadening Perspectives (BP), and Consolidation}---and introduced three additional types of cues aligned with the identified indicators of critical thinking: \textit{Source Engagement} (SE), \textit{Persistent Inquiry} (PI), and \textit{Independent Thinking} (IT). The Orienting and Monitoring cues, which help establish evaluative criteria for GenAI responses and prompt comparison with prior knowledge, respectively, were delivered at fixed intervals. The remaining cues were dynamically triggered based on user interactions. These cues, described in detail below, serve the following purposes: The PI cue encourages follow-up questions in pursuit of depth of understanding, the SE cue promotes active engagement with sources cited in the AI responses, the IT cue stimulates reflection and synthesis through note-taking, and the BP cue encourages consideration of unexplored perspectives.
The SE, PI, and IT cues are instantiated in two variants: a \textit{regular} variant that encourages under-exhibited desirable behaviors (e.g., ``Are there parts of the AI response for which you need more details or evidence Consider going through the linked source...") and a \textit{reinforcement} variant that acknowledges and strengthens desirable behaviors that are already demonstrated (e.g., ``Great job engaging with the sources! This is helpful for going beyond surface level understanding."). This design choice was informed by feedback from pilot studies, in which participants expressed a preference for cues that recognized ongoing effective behaviors rather than redundantly prompting actions they believed they were already performing. 
While we prompt GPT-5\cref{main_footnote} to determine which of these variants to deliver, the actual cue messages are predefined\cref{main_footnote}, for consistency and preserving the integrity of metacognitive support---guiding users’ thinking and search behavior without offering explicit search recommendations \cite{bannert2013scaffolding}. We explored dynamically generating cue messages via LLM prompting, but this approach produced noisy outputs that did not consistently meet established criteria for effective metacognitive scaffolding. Future work may investigate more advanced prompting strategies to support reliable dynamic cue generation.

MetaCues currently delivers cues at predetermined intervals in a fixed sequence: an Orienting cue at session start, a Monitoring cue after the first query, followed by SE → IT → PI → BP. These four dynamic cues cycle in this order until the session ends. The initial SE cue is delivered 3-minutes after the session begins, and subsequent cues are triggered at 2.5-minute intervals. Future work may explore more adaptive cue scheduling strategies, including dynamically adjusting cue timing and gradually fading cues as users internalize desirable behaviors.

The regular variant of the \textbf{SE cue} is sent if any AI response containing sources has zero source clicks, else its reinforcement variant is sent.
If there are no sources in any of the responses so far, a special message is sent to encourage engagement with sources when they do appear.
For the \textbf{PI cue}, MetaCues identifies if a user is asking relevant follow-up questions by prompting GPT with the chat history and search topic along with positive and negative examples of relevant follow-up queries. 
The regular variant is sent if it is determined that the user has not asked any relevant follow-up questions thus far, otherwise the reinforcement variant is sent.
For the \textbf{IT cue}, MetaCues prompts GPT to compare the user's notes with AI responses and content scraped from the sources in these responses, to determine if the notes contain novel viewpoints, such as questions the user may have about the information obtained from searching. If no novel viewpoints are found, the regular variant is sent, else the reinforcement variant is sent.
If notes are empty, a special message is displayed to encourage the user to take notes and reflect on their prior knowledge and note any unanswered questions.
Finally, the \textbf{BP cue}, which promotes exploration of overlooked perspectives, does not include a reinforcement variant, as achieving comprehensive exploration within the brief study duration is unlikely. 

Once a cue is triggered, it is queued to be displayed in an activity-aware process. MetaCues waits for natural pauses in user activity (3-second idle), and displays a new cue only when the interface is visible to the user, and there has been some recent user activity within the last 5 minutes.
In case such an opportunity is not found, the cue is shown 60 seconds after generation. 
This approach minimizes distraction while ensuring that the cues are not missed. When a new cue is displayed, a pulsing glow around the cue icon draws attention. The pulsing effect stops when the user acknowledges the cue by clicking the thumbs-up button next to it.

\section{Study Design}
We conducted an online between-subjects factorial experiment comparing (\textit{MetaCues}) against a (\textit{Baseline}) tool without metacognitive cues, across two information-seeking tasks. 
The Baseline tool consists of the chat box and notepad with identical functionality as MetaCues, but does not include the Cues box. 
The two search topics were selected to elicit participant interest---albeit at varying levels---based on four primary criteria. Specifically, the topics needed to be: (i) timely, in order to foster authentic motivation to learn and ensure a baseline level of participant familiarity; (ii) open-ended, to encourage the exploration of diverse perspectives; (iii) readily understandable, to ensure participant engagement; and (iv) conducive to informed judgment, necessitating the consideration of multiple viewpoints. Additionally, the first topic was intentionally chosen to be more controversial than the second, enabling an evaluation of the impact of MetaCues across topics with differing levels of controversy. Accordingly, we selected the following two topics:
\begin{itemize}
    \item \textbf{Social Media}: ``Considering how different social media platforms impact mental health in teenagers, should social media use be banned for individuals below the age of 16?"
    \item  \textbf{Music}: ``Given the cognitive and physiological effects of listening to music while studying or test-taking, should students be allowed to listen to music during school exams?"
\end{itemize}

Participants were randomly assigned to one of four topic-condition groups: \textit{Baseline-Social Media}, \textit{MetaCues-Social Media}, \textit{Baseline–Music}, or \textit{MetaCues-Music}.

\subsubsection*{\textbf{Study Procedure}}

After providing consent, participants answered demographic and LLM usage questions.
Next, they were introduced to their assigned tool through a brief tutorial, and prompted to conduct research on the assigned topic while taking notes as if to prepare for writing an essay, using only the assigned tool. Participants were encouraged to gather evidence from reliable sources, understand the topic from multiple perspectives, and explore sources linked in the AI responses.
Before starting the task, participants rated their familiarity with and interest in the topic on a 5-point Likert scale and were informed they had up to 25 minutes to complete the task. A timer was displayed in the chat panel, and participants could end the session early, otherwise it ended automatically after 25 minutes. 

\subsubsection*{\textbf{Data on Attitudinal Judgments}}
After completing the task, participants rated their attitudinal judgments regarding the topic and their level of confidence in their judgment, each on a 5-point Likert scale. For the social media topic, participants rated their agreement with banning its use in schools; for the music topic, they rated their agreement with allowing its use during school exams. 

\subsubsection*{\textbf{Data on Search Behaviors}}
We computed the following behavioral measures from system logs: 
(1) \textit{search duration}, 
(2) \textit{time spent outside the interface}, 
(3) \textit{total typing time}, 
(4) \textit{number of queries}, 
(5) \textit{average words per query}, 
(6) \textit{number of sources clicked}, 
(7) \textit{click-through rate}, and 
(8) \textit{query divergence}.
Measure (2) estimates the time participants spent on the linked sources, which opened in a new tab. Measure (7) captures the ratio of unique sources clicked to the number of all unique sources linked in all AI responses during the session.
Measure (8) quantifies how semantically divergent a user's queries are within a given topic: lower divergence indicates more focused querying, whereas higher divergence reflects broader conceptual exploration. To measure query divergence, we first trained vectorized embedding representations separately for each topic. Each query was represented as a 384-dimensional embedding generated by the \textit{all-MiniLM-L6-v2} 
sentence-transformer model \cite{reimers2019sentencebert}. Further, queries were L2-normalized and cosine distance between two embeddings was used to capture semantic dissimilarity. For a user $u$, we computed a centroid of their queries' embeddings: $\mathbf{c}_u = (1/n_u)\sum_i \mathbf{v}_i$. Then, we measured the cosine distance between each query $\mathbf{v}_i$ and its centroid as $d_i = 1 - (\mathbf{v}_i \cdot \mathbf{c}_u)/(\|\mathbf{v}_i\|\|\mathbf{c}_u\|)$. Query divergence is the mean of these distances $\bar{D}_u = (1/n_u)\sum_i d_i$.
\\
\newline In addition to this data, we evaluated participants' learning outcomes through a post-test, and captured artifacts reflecting their engagement during the information-seeking process, including notes, end-of-search summaries, and conversations with the assigned AI. The analysis of this data is reserved for future work.

\section{Results}

\subsubsection*{\textbf{Participants Overview}}
Participants were recruited via Prolific\footnote{\href{https://www.prolific.com}{https://www.prolific.com}}
and compensated at \$15 per hour. Eligible participants were U.S.-based individuals currently enrolled in or holding a college degree. The study lasted approximately one hour. 
175 participants completed the study, of which 146 passed all attention checks and constituted the study sample. Ages ranged from 18–59 (median: 18–24); 73 identified as male, 67 as female, 6 as non-binary. 
Kruskal-Wallis H tests
revealed no significant differences between the four topic-condition groups in LLM use frequency ($H(3)=1.08$, $p=0.78$) and frequency of using LLMs for search purposes ($H(3)=0.152$, $p=0.99$).

\subsubsection*{\textbf{Topic Familiarity and Interest}}
Mann-Whitney U tests revealed a significant difference in participants' perceived familiarity with the search topics ($U = 3459.50$, $p = 0.001$), with participants reporting more familiarity with the social media topic ($M=2.85$, $SD=0.88$) than the music topic ($M=2.36$, $SD=1.08$). There was no significant difference in their level of interest between the social media ($M=3.51$, $SD=1.04$) and music ($M=3.43$, $SD=1.18$) topics ($U = 2737.00$, $p = 0.770$). 

\subsubsection*{\textbf{Effects on Search Behaviors \& Confidence in Attitudinal Judgments}}

\begin{table*}[h]
\centering
\begin{tabular}{lcccc}
\toprule
\multirow{2}{*}{\textbf{Measure}} & \multicolumn{2}{c}{\textbf{Social Media}} & \multicolumn{2}{c}{\textbf{Music}} \\
\cmidrule(lr){2-3} \cmidrule(lr){4-5}
 & \textbf{Baseline: M (SD)} & \textbf{MetaCues: M (SD)} & \textbf{Baseline: M (SD)} & \textbf{MetaCues: M (SD)} \\
\midrule
Search duration (s) & 1299.27 (386.48) & 1241.93 (365.43) & 1186.39 (400.76) & 1284.83 (405.38) \\
Time spent outside interface (s) & 150.79 (150.35) & 370.01 (1103.84) & 206.14 (217.77) & 248.94 (303.02) \\
Total typing time (s) & 353.09 (213.67) & 307.67 (222.27) & 272.03 (143.03) & 285.81 (173.26) \\
Number of queries & 7.39 (6.76) & 7.35 (5.67) & 5.95 (3.75) & 9.25 (7.71) \\
Average words per query & 15.76 (6.60) & 21.40 (19.45) & 16.12 (12.83) & 14.65 (6.62) \\
Average words per AI response & 280.06 (112.87) & 264.24 (80.97) & 210.12 (76.79) & 218.60 (54.62) \\
Number of sources clicked & 2.61 (2.94) & 2.76 (3.16) & 2.49 (2.71) & 2.78 (2.52) \\
Click-through rate & 0.21 (0.24) & 0.24 (0.26) & 0.23 (0.28) & 0.27 (0.27) \\
Query divergence & 0.23 (0.17) & 0.26 (0.15) & 0.28 (0.14) & 0.34 (0.15) \\
\bottomrule
\end{tabular}
\caption{Descriptive statistics for each search behavior measure.}
\label{tab:descriptive-stats}
\end{table*}

Table \ref{tab:descriptive-stats} reports the mean and standard deviation for each search behavior measure across conditions and topics. To assess the effects of searching with versus without cues, we fit Generalized Linear Models (GLMs) with appropriate link functions as the data did not meet normality assumptions. Condition, topic, and their interaction were used as fixed effects. 
We used a negative binomial (log link) for count data, gamma (log link) for time-based measures and click-through rate, and Gaussian (identity link) for average words per query. 
For query divergence, we ran separate Mann–Whitney U tests per topic as we trained distinct embedding models for each topic. 
Lastly, for attitudinal judgments, we conducted a two-way ANOVA with topic, condition, and their interaction as the independent variables.

\begin{figure}[]
  \centering
  \setlength{\abovecaptionskip}{-2pt}   
  \setlength{\belowcaptionskip}{-10pt}  
  \begin{subfigure}{0.4\linewidth}
    \includegraphics[width=\linewidth, trim=15 10 23 7, clip]{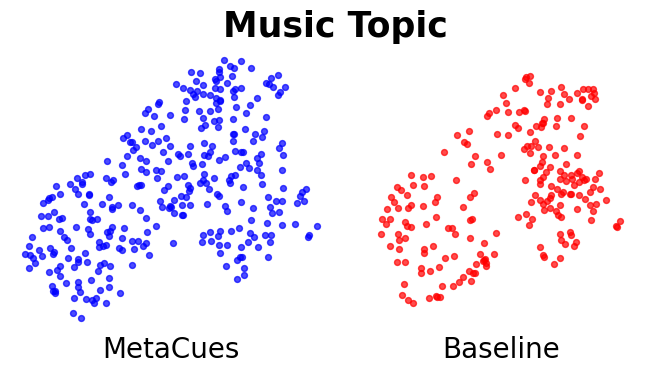}
    \label{fig:umap_music}
  \end{subfigure}
  \hfill
  \begin{subfigure}{0.4\linewidth}
    \includegraphics[width=\linewidth, trim=23 10 15 7, clip]{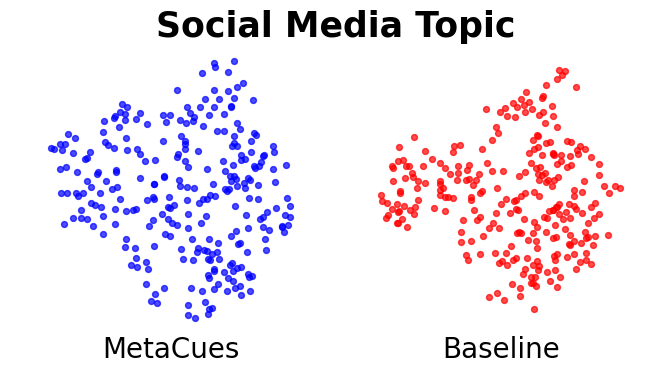}
    \label{fig:umap_social}
  \end{subfigure}
  \caption{UMAP visualizations of query embeddings for music (left) and social media (right) search topics.}
  \label{fig:umap_topics}
\end{figure}

We now report the results of the statistical analyses.  
For the music topic, we found that users' query divergence in the MetaCues condition ($M=0.34$, $SD=0.15$) was significantly higher ($U=511.5$, $p=0.045$, $d=0.40$) than Baseline ($M=0.28$, $SD=0.14$). 
For the social media topic, query divergence for MetaCues ($M=0.26$, $SD=0.15$) was slightly higher than Baseline ($M=0.23$, $SD=0.17$), but the difference was not statistically significant ($U=610.5$, $p=0.27$, $d=0.16$). 
Additionally, to explore how queries differed semantically across conditions, we visualized their embeddings using UMAP \cite{mcinnes2018umap}, projected into a two-dimensional latent space that preserved relative semantic distances. Figure~\ref{fig:umap_topics} shows the Baseline and MetaCues groups side by side for both topics. For the music topic, the MetaCues group appears more widely dispersed, extending into regions that are sparse for the Baseline group, suggesting broader and more exploratory query formulation. For social media, the two groups show similar overall density, though the MetaCues group exhibits a slightly broader outer contour. These visualizations are consistent with the results of the statistical analyses reported above.

For both time spent outside the interface (likely on linked sources) and click-through rate, participants in the MetaCues condition showed higher means across topics (see Table~\ref{tab:descriptive-stats}), though these differences were not significant.
Turning to the remaining search behavior measures, mean time spent outside the interface and click-through rates were higher in the MetaCues condition, but neither difference was statistically significant. For the number of queries, the MetaCues condition showed a notably higher mean for the music topic, but this effect was also non-significant. No significant main effects of condition, topic, or their interaction were observed for overall search duration, total typing time, average words per query, and number of sources clicked. Detailed statistical analyses results can be found here\cref{main_footnote}.

Regarding confidence in attitudinal judgments, a two-way ANOVA revealed a significant main effect of condition, ($F(1, 142)=4.53$, $p=0.035$), indicating that participants in the \textit{MetaCues} condition reported higher confidence than those in \textit{Baseline}. There was no significant main effect of topic ($F(1, 142)=0.03$, $p=0.857$), and no significant interaction between condition and topic ($F(1, 142)=0.40$, $p=0.530$). 

\section{Discussion}
This study demonstrates the feasibility of automatically generating metacognitive cues to guide GenAI-based information seeking, verifying the findings from Singh et al. \cite{singh2025enhancing}.  
However, the effects of MetaCues varied by search topic. The finding that MetaCues led to significantly greater query divergence for the music topic---on which participants reported lower prior familiarity compared to the social media topic, and which was less controversial in nature---suggests that MetaCues may be more effective in supporting broad exploration for topics with which users have lower perceived familiarity and less strongly held opinions. In contrast, the social media topic was widely discussed at the time of the study, which may have contributed to participants holding more established or polarized views. As a result, MetaCues may not have exerted as strong an influence on search behavior for this topic. Future work involving qualitative and in-depth analyses of participants' search interactions and post-test outcomes is needed to provide further insight into how MetaCues shaped learning across both topics, and how these effects were mediated by participants’ familiarity with and the controversial nature of the topics.

Notably, MetaCues led to significantly higher confidence in the resulting attitudinal judgments compared to the baseline tool, which reflects potentially greater perceived epistemic grounding resulting from deeper engagement with sources, reflection, and inquiry. This suggests that metacognitive cues are effective for aiding the consolidation of understanding necessary for judgment formation \cite{reyna2003memory}. 
However, level of confidence is not reflective of the actual impact on their learning outcomes. Therefore, future work should further examine whether such confidence is well-calibrated, how it evolves over longer-term or higher-stakes tasks, and also conduct qualitative exploration of participant interactions and inquiry patterns.

\section{Conclusion}
We presented MetaCues, an interactive system that automatically generates metacognitive cues to support GenAI-based search. In an online between-subjects study, we found that it promotes more active and diverse inquiry than a baseline system without cues, particularly for topics that are less controversial and with which users are less familiar. Further, it leads to higher confidence in resulting attitudinal judgements. 
However, the study's modest sample size ($N=146$) limits statistical power; future work with larger samples could further examine how metacognitive cues affect different types of search tasks. As cue generation in MetaCues was timed according to study constraints, subsequent research should explore more adaptive cue delivery and the gradual fading of metacognitive support as users gain experience. Future work should also include qualitative analyses of participant interactions and inquiry patterns.

\bibliographystyle{ACM-Reference-Format}
\bibliography{REFERENCES}

\end{document}